# Understanding the Factors Influencing Self-Managed Enterprises of Crowdworkers: A Comprehensive Review

Alexandre Prestes Uchoa[1] [a] and Daniel Schneider[1,2] [b]
[1]*Postgraduate Program in Informatics, PPGI/UFRJ, Rio de Janeiro, Brazil*
[2]*Tércio Pacitti Institute of Computer Applications and Research, NCE/UFRJ, Rio de Janeiro, Brazil*
*alex.uchoa@gmail.com, schneider@nce.ufrj.br*

Abstract: This paper investigates the shift in crowdsourcing towards self-managed enterprises of crowdworkers (SMECs), diverging from traditional platform-controlled models. It reviews the literature to understand the foundational aspects of this shift, focusing on identifying key factors that may explain the rise of SMECs, particularly concerning power dynamics and tensions between Online Labor Platforms (OLPs) and crowdworkers. The study aims to guide future research and inform policy and platform development, emphasizing the importance of fair labor practices in this evolving landscape.

## 1 INTRODUCTION

The crowdsourcing landscape is undergoing a significant transformation, redefining how collective intelligence is utilized in contemporary settings. Crowdsourcing has evolved from a universally accessible model to a sophisticated ecosystem of platforms that selectively bridge specific segments of the crowd with recruiters, thereby mediating these interactions (Kittur et al., 2013; Zhao and Zhu, 2014; Lopez, Vukovic, and Laredo, 2010).

As these platforms began addressing more complex tasks, they encountered challenges in assembling teams with the necessary expertise. This process can be both resource-intensive and costly, particularly in fluctuating market conditions (Ho and Vaughan, 2012). The intricacy of these tasks and constrained platform oversight necessitated enhanced collaboration among workers and the provision of greater autonomy and creative freedom (Lykourentzou et al., 2019), prompting the need for innovative management strategies.

At the same time, "coming from the other side of the fence", a particularly noteworthy development in this domain is the emergence of self-organized groups of crowdworkers who independently manage and execute complex macrotasks. These groups represent a paradigm shift from the traditional, platform-centric workforce model to a self-directed, enterprise-like collaboration (Wang et al., 2020; Huo, Zheng, and Tu, 2017), thus challenging the direct control exerted by platforms. SMECs represent a potential paradigm shift, offering crowdworkers greater autonomy and potentially improved working conditions compared to traditional OLP work.

However, despite its promise, SMEC is still an emerging and localized phenomenon with scant scholarly attention thus far (Wang et al., 2019; 2020; 2021; 2023). To bridge this research gap, we have thoroughly reviewed the extant literature, concentrating on specific factors and challenges associated with macrotask platform work that are pivotal in defining and differentiating this nascent model from traditional platform work. We aim to shed light on this area, thus providing crucial insights into the possible origins and developmental trajectories of SMECs and informing subsequent research directions.

The rest of the paper is organized as follows. In the next section (Background), we provide some context on the main topics of this study. Next, a detailed description of the methodology is presented in Section 3, followed by the results (Section 4), and an analysis of the results with a discussion of the main issues raised (Section 5). Finally, we depict our limitations (Section 6), future research directions (Section 7), and conclusions (Section 8).

## 2 BACKGROUND

Paid crowdsourcing is a type of socio-technical work characterized by a triangular relationship between recruiters (companies and individuals), crowds, and platforms (Kittur et al., 2013).

[a]  https://orcid.org/0000-0002-2028-0252
[b]  https://orcid.org/0000-0003-2987-4732

Platforms are the ones that bring recruiters and the crowd together (Zhao and Zhu, 2014), thus intermediating the interactions and communications between the other two (Lopez, Vukovic and Laredo, 2010). Nowadays, there is a panoply of different types of crowdsourcing, ranging from corporate to social and public contexts (Vianna, Peinado and Graeml, 2019), and a proliferation of for-profit crowdsourcing platforms that, in effect, determine who can actually participate. According to Chaves et al. (2019), different forms of public participation and engagement can be achieved in such platforms that harness crowd workers.

This kind of paid crowdsourcing, also called online work, digital work, or online labor, has become widely recognized for its effectiveness in distributing not complex tasks to a large number of individuals in the crowd because they do not require specific skills and can thus be performed quickly and repetitively. Such tasks are typically outsourced by OLPs like Amazon Mechanical Turk, which pays very little for them (De Stefano, 2016; Deng, Joshi and Galliers, 2016).

On the other side of the spectrum of task complexity and size are the so-called macrotasks. These are difficult tasks that are sometimes even impossible to break down into smaller, easier subtasks (Robert, 2019). For this reason, unlike their simpler cousins, macrotasks usually require specific skills and knowledge to be accomplished (Schmitz and Lykourentzou, 2018). There is a wide variety of problems and types of tasks that today are being addressed with the help of macrotask crowdsourcing (Wang et al., 2021; Gimpel et al., 2023; Kohler and Chesbrough, 2019; Mcgahan et al., 2021; Geiger and Schader, 2014).

OLPs, in general, as an intermediate, act facilitating functions such as task and contract management and dispute resolution (King, 1983; Shafiei Goal, Avital and Stein, 2019), but also deciding whether and how to divide these complex tasks into smaller subtasks, distributing them to workers according to their assessment of their skills and capabilities, managing them and their interdependencies (Kittur et al., 2013). The distinction between both types of tasks is crucial as it affects how OLPs handle them (Cheng et al., 2015) and how they shape their work processes accordingly (Leimeister et al., 2016).

Macrotask crowdsourcing may be considered to be a manifestation of post-bureaucratic work (Barley and Kunda, 2001; Seppänen et al., 2021), in which expertise is distributed outside bureaucracies, organizations, and hierarchies. Kittur et al. (2013) see such a form of work as a new form of technological Taylorism in which the rules and guarantees of subordinate work do not apply. Because it is not feasible to use traditional control mechanisms, OLPs safeguard themselves in various ways. According to Kornberger, Pflueger and Mouritsen (2017), macrotask OLPs operate like "evaluative infrastructures that create competition and incentives out of the differences among workers and establish power through the decentralization of control". One way to minimize the efforts needed to control workers in a macrotask setup is to ensure that their aspirations align with the OLP's goals (Schörpf et al., 2017). In pursuing this goal, some OLPs adopt hiring processes comparable to traditional employers by conducting background checks, face-to-face interviews, skills assessments, and even test drives (Kuhn and Maleki, 2017).

Another key component of macrotasking is collaboration among workers (Kittur et al., 2013; Gimpel et al., 2023; Lykourentzou et al., 2019). On the other hand, such teams require increased coordination effort (Kerr and Tindale, 2004; Kittur and Kraut, 2008). Using the very interdependence of the team and employing managerial practices that promote autonomy has shown beneficial effects (Gagné and Deci, 2005). According to Kerr and Tindale (2004), the quality of a group can be measured by the ability of its members to reach an agreement about what is to be done. Yet, giving workers more autonomy, creative freedom, and initiative requires the OLPs to innovate in how they manage the crowd (Lykourentzou et al., 2019).

In this scenario, an intriguing development emerges, which is that of independent, self-constituted, and managed teams of crowdworkers, in distinction from the groups constituted and coordinated by the OLPs. Although still incipient in the West, these self-organized teams, or companies, called by Wang et al. (2020) "crowdfarms", is a phenomenon mainly observed in China, a country with a mature crowdsourcing market where, in 2017, there were already 30 million crowdworkers serving more than 190,000 companies and individuals from all over the world and generating a total turnover of approx. $700 million (Huo, Zheng and Tu, 2017).

What was once a market dominated by individual workers seeking extra revenue in their spare time by executing microtasks has seen their gradual replacement by small organizations that perform crowdsourcing tasks en masse (Wang et al., 2019), employ full-time salaried workers, and operate in formal (i.e., physical) workplaces such as business offices. The ZBJ platform, for instance, one of the most prominent crowdsourcing OLP in China with more than 19 million active crowdworkers, acts as a kind of "incubator" for these crowdsourcing companies, having already supported the creation of more than 150,000 of them since 2016 by providing them with services, such as financial and legal, as well as physical workspaces in 26 major cities in China, calling them "crowdsourcing factories" (Wang et al., 2021). The authors attribute the emergence of these organizations in China to the gradual transformation and increased complexity of tasks posted on OLPs, coupled with favorable government policies, such as the "mass entrepreneurship and innovation program", as well as support from the very OLPs, including the aforementioned ZBJ factories.

Unlike "flash organizations" (Valentine et al., 2017), where random solo workers are automatically organized into a hierarchy according to their abilities in a temporary structure computationally constructed to handle a specific complex task, in these SMECs, the decisions are all up to the very company, including the breakdown of complex tasks. Thanks to the extensive experience in crowdwork and the mastery that the managers of these companies usually have of their work arrangement and capabilities, they can procure, decompose, and allocate tasks internally, designing their own workflows by adapting them to their teams and thus being more effective.

For the workers at these SMECs, the personalized workflows improve their understanding of their personal duties and roles and make things easier by setting the standards for cooperation (Wang et al., 2023).

Workers are also attracted by the better payments complex tasks offer, by the guarantees secured through legal contracts, and by the possibility that these companies provide for establishing interpersonal relationships with customers, which leads to more business opportunities (Wang et al., 2020; 2021). The experiences and context in these self-managed companies end up underpinning their motivations, the ways they engage with crowdworking, the tasks they work on, and the OLPs they use.

However, despite the valuable empirical insights provided by Wang et al. (2019; 2020; 2021; 2022), much is still unknown about crowdwork enterprises like the crowdfarms and their workers. While SMECs, much like OLP-controlled crowdworkers, also have to deal with aspects such as problematic requirements and specifications, deadlines and costs, customer acceptance criteria, prospecting and reputation, OLP policies and algorithms, payments and defaults, they do it in different scales and manners.

What exact factors and work practices do these SMECs employ, what other forms of SMECs besides crowdfarms can be thought of, and what barriers hinder their emergence in other markets are just some examples of what must be further explored. SMECs represent an expansion of the traditional understanding of paid crowdworking and reflect the ongoing evolution of the field. And as with any new socio-technical advance, it reveals new challenges and potential spaces for investigation.

This review pinpointed eight principal areas of contention in macrotask platform work from the workers' perspective, utilizing them to delve into and refine our grasp of the underlying dynamics of this emerging model. These areas encompass payment schemes, trust and reputation systems, control and autonomy, exploitation and unfair treatment, demands for improved conditions, algorithmic bias, crowdworker unity, and empowerment. By adopting a socio-technical lens, the review methodically explores this broad array of topics, highlighting the intricate interplay of power, agency, and mutual dependence that characterizes the relationship between workers and OLPs.

## 3 METHODOLOGY

Our methodology for this review centered on examining existing research on aspects and factors that may underpin the emergence of SMECs. We aimed to answer the following three key review questions by identifying studies and assessing to what degree they address the unique aspects that characterize this phenomenon.

RQ1) What are some commonly overlooked aspects of crowdworking that characterize SMECs?
RQ2) What dynamics and sources of tension between crowdworkers and OLPs have been investigated?
RQ3) What potential areas for further research can be identified?

### 3.1 Search strategy

Early in the first exploratory searches for literature, some important difficulties were already noticed.

The results were remarkably insignificant or diffuse when searching different databases using expressions such as "groups" or "self-managed teams of crowdworkers" as proxies for SMECs. Research articles on this phenomenon are still rare, and the few that have been found have their authorship concentrated in a few researchers (Lykourentzou, Robert JR and Barlatier, 2022; Wang et al., 2019; 2020; 2021; 2023), probably due to the youth and locality of the phenomenon. The lack of agreement on a common denomination and definition for SMECs and an objective, precise, and widely accepted definition of macrotask and complex crowdwork also hindered the first searches.

With this in mind, the review decided to design the search so that it could retrieve a broader spectrum of studies that include subjects and aspects of macrotask crowdworking that are, we believe, at the core of what most clearly distinguish SMECs from typical OLP-controlled forms of crowdwork. The four aspects of crowdworking that the review chose were:

A1. Workflow, task decomposition, coordination;

A2. Worker selection, assignment, incentives;

A3. Team composition;

A4. Power dynamics, precarity, sources of tension.

## 3.2 Sources and search process

To systematically select relevant literature, we employed a multi-step search strategy using the two major indexing databases: Scopus and Web of Science (WoS).

1. For every search instruction tested, it was ensured that selected articles considered referential in the field of interest of the review were among the search results in at least one of the databases used.
2. For each search result, the keywords of the retrieved articles were analyzed in search of relevant new words to search for.
3. As the process progressed, new referential articles were added, and the new improved search sentences had to retrieve them as well.
4. Of the successful alternatives, the one capable of retrieving the smallest set of results and still containing all the referential articles was chosen.

Figure 1 presents the final version of the search instruction used, depicting the combination of terms and logical operators used in article searches in the chosen indexing bases.

---

*(crowdwork OR crowd work OR crowdsourcing OR crowdsource  
OR crowdfarm OR crowd farm OR self-assembled crowd  
OR online labour platform OR online labor platform  
OR digital labour market OR digital labor market  
OR online labour market OR online labor market)  
AND  
(platform work OR teamwork OR project team OR collective labor OR co-creation  
OR knowledge task OR knowledge work  
OR expert domain knowledge OR expert knowledge OR expert task  
OR interdisciplinary project OR interdisciplinary task OR interdisciplinary work  
OR creative collaboration OR creative work OR creative task  
OR crowd-based workflow OR crowdbased workflow OR crowdsourced workflow  
OR macrotask OR macro-task OR macro task OR macrotasking  
OR complex task OR non-decomposable task)*

---

Figure 1: Combination of terms and logical operators used in searches for articles in the Scopus and Web of Science indexing databases.

## 3.3 Inclusion and Exclusion Criteria

Our inclusion criteria were specific to peer-reviewed, not redundant journal and conference articles published within the past decade (criterion C1), focusing on:

I. Paid crowdworkers;
II. Online Labor Platforms (OLPs);
III. Macro, complex, creative, knowledge-based, interdisciplinary, non-decomposable tasks;
IV. Decomposition or workflows of tasks;
V. Collaboration or co-creation.

For this reason, all other types of outputs were excluded, together with the following cases:

- Studies addressing impacts of COVID pandemic;
- Reviews, workshops, tutorials, thesis & dissertations;
- Articles with no abstract available;
- Article's full text not available in English;
- Article's full text not available for reading at the time of the writing.

We also deliberately excluded studies that either use crowdsourcing as a subsidiary means to accomplish their objectives (such as obtaining data), propose design principles for new crowdsourcing endeavors, or address types of crowdsourcing that are fundamentally distinct from the paid work for OLPs (criterion 2), among them:

- Voluntary or unpaid crowdworking (e.g., citizen science, contests, games);
- Sharing economy, online communities, open-source project, collaborative crowdsourcing;
- Mobile, geographic, collocated or spatial crowdsourcing;
- Crowdsourcing to customers or consumers;

- Enterprise, marketing, industry crowdsourcing;
- Crowdsourcing by the government;
- Passive crowdsourcing (e.g., IoT, body sensors for active or passive data collection);
- Crowdfunding.

Exceptions would be tolerated when contributions were deemed relevant and extendable to crowdwork on macrotasks. Finally, the selection should also exclude propositional studies with contributions in the form of technological solutions, new approaches and methods, particularly those aimed at a single specific party. This includes those aimed at OLPs and requesters with propositions such as boosting worker productivity and task quality or reducing costs and risks (*criterion 3*).

This approach was chosen, firstly, because a wider scope of audience and subject often encapsulates a more diverse range of perspectives and insights. Secondly, such papers tend to foster interdisciplinary dialogue and broader applicability, extending their relevance beyond a single, specialized domain or problem, which is especially valuable when exploring the realm of SMECs.

The implementation of a protocol for assessing the quality of the retrieved studies was waived by the review process. This decision was rooted in the belief that imposing a quality protocol at this juncture of such a nascent field could inadvertently narrow the scope of our review, thus omitting valuable insights and emerging lines of inquiry.

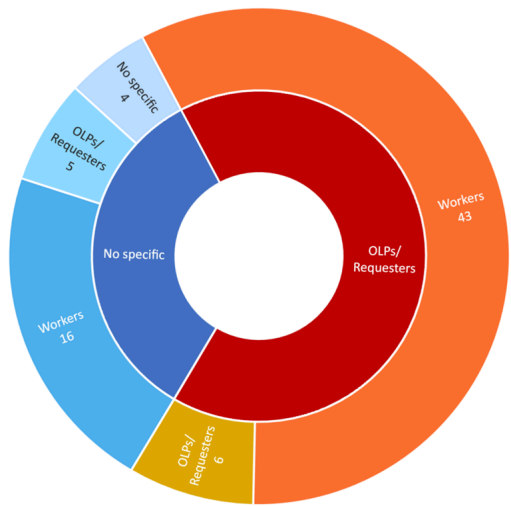

Figure 2. Two-level sunburst chart of articles before the application of selection criterion 3. The inner ring categorizes articles by their intended audience/aim. The outer ring details the main papers studied.

## 4 RESULTS

From an initial pool of 1003 publications (313 from WoS and 687 from Scopus), we downloaded all metadata, reduced 62 redundancies, and retained recent peer-reviewed journal articles. Utilizing OpenAI's GPT-3.5 Turbo, we programmatically analyzed the abstracts of 713 remaining articles, generating seven syntheses to distill their key points.

Seven syntheses were generated, addressing key questions about each study's purpose, contributions, evidence, central argument, type of crowdsourcing, results, and data used. These syntheses streamlined the application of thematic inclusion and exclusion criteria (criterion 1 & 2), aiding in identifying studies irrelevant to our review. This process narrowed the field to 74 articles. GPT-3.5 Turbo was again employed, this time analyzing the full texts of these articles, further assisting in our research synthesis. In order to apply criterion 3, we categorized these 74 articles according to three dimensions:

Audience/End User: Identifies the primary target group for the study's contributions, particularly when directed towards a specific audience.
Main Party Studied: the party that primarily benefits from the solutions and approaches proposed in the study (OLPs/requesters, crowdworkers, no specific).
Aspect Studied: The predominant aspect of crowdworking that the study's contributions address (A1 to A3).

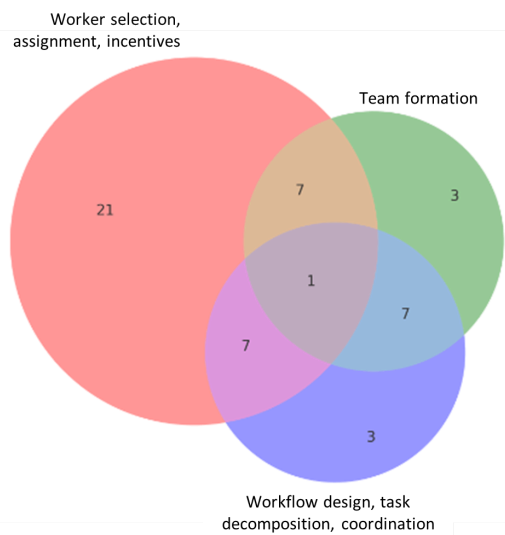

Figure 3. Distribution of the articles aimed at platforms/requesters that address one or more of the three aspects of crowdworking we choose to focus on: A1. Worker selection, task assignment, incentives; A2. Team formation; and A3. Workflow design, task decomposition, coordination. Overlapping regions indicate articles addressing multiple aspects.

LLM-assisted categorization is hampered by their sensitivity to task wording. Mitigating ambiguity and vagueness, especially in category names, is crucial for reliable and reproducible categorization. Figures 2 and 3 illustrate the outcomes of this categorization process. The initial set of 74 articles primarily offered solutions and propositions for platforms and/or requesters, focusing on worker selection, task assignment, and incentive mechanisms (Figure 2).

The review's selection phase concluded by narrowing down to 25 articles that target a more general audience, as illustrated in Figure 4. This choice was influenced by the fact that papers with a broader audience typically adopt an analytical, observational, or critical approach, aligning with our review's objectives to provide comprehensive insights into the thematic exploration of crowdworking dynamics.

The remaining 49 articles, focused on specific solutions for platforms and requesters, were excluded due to their narrower scope. This strategic decision was based on our criteria to include studies with broader relevance, ensuring a more universally applicable understanding of the subject.

## 5 ANALYSIS AND DISCUSSION

OLPs play a pivotal role in shaping task availability, payment, and working conditions while relying on skilled crowdworkers, whose needs are vital for both task completion and platform success. This interplay directly affects crowdworkers' autonomy and satisfaction and is crucial in the context of SMECs, as it influences their operation and effectiveness. Given the variability across different OLPs and worker groups, this area is a rich vein for research. To delve deeper into these nuances, we have broken down this aspect (A4) into eight key factors, each shedding light on the complexities of this ecosystem, as follows:

- *F1. Payment schemes*: the impact of payment schemes on the relationship between crowdworkers and OLPs;
- *F2. Trust and Reputation Systems*: the role of trust and reputation systems in shaping the relationship between crowdworkers and OLPs;
- *F3. Control and autonomy*: the impact of autonomy, freedom and control over tasks;
- *F4. Exploitation and unfair treatment*: potential for exploitation or unfair treatment of crowdworkers by OLPs;
- *F5. Demand for better conditions*: collective action or organizing among crowdworkers to advocate for better working conditions or autonomy;
- *F6. Algorithmic bias*: potential for algorithmic bias or discrimination in the OLP decision-making processes;
- *F7. Crowdworker Unity*: crowdworker solidarity or collaboration in response to OLP practices;
- *F8. Crowdworker empowerment*: crowdworker empowerment or agency in shaping the OLP policies and practices.

We focused on determining the presence or absence in the 25 articles including discussions related to these factors. This approach allowed us to quantitatively assess the extent to which these underlying factors were considered in the body of literature at hand. To achieve this, we categorized each paper based on whether it explicitly mentioned or engaged with each of our underlying factors. It's important to note that this analysis was binary in nature; we marked a factor as 'addressed' if it was either discussed in detail or merely pointed out in the paper as an important aspect to be considered.

This method provided a broad overview of the thematic landscape of the field, indicating which areas have been given more or less attention in academic discourse. However, it does not assess the depth or quality of the coverage for each topic within the individual papers. Our goal is to map the prevalence of certain themes and identify potential gaps in the literature rather than to perform a qualitative analysis of the discussions surrounding these themes. Consequently, the results offer insights into the frequency of topic appearances across our set of papers, reflecting trends and potential areas for further research in the realm of crowdwork and OLPs.

| Article | F1 | F2 | F3 | F4 | F5 | F6 | F7 | F8 |
|---|---|---|---|---|---|---|---|---|
| (ALACOVSKA; BUCHER; FIESELER, 2022) | | | | | | | | |
| (ARCIDIACONO; BORGHI; CIARINI, 2019) | | | | | | | | |
| (BRAESEMANN et al., 2022) | | | | | | | | |
| (BRAESEMANN; LEHDONVIRTA; KÄSSI, 2022) | | | | | | | | |
| (GERBER, 2022) | | | | | | | | |
| (GIMPEL et al., 2023) | | | | | | | | |
| (GLAVIN; BIERMAN; SCHIEMAN, 2021) | | | | | | | | |
| (GOL; AVITAL; STEIN, 2019) | | | | | | | | |
| (GOL; STEIN; AVITAL, 2018) | | | | | | | | |
| (HULIKAL MURALIDHAR; RINTEL; SURI, 2022) | | | | | | | | |
| (IMMONEN, 2023) | | | | | | | | |
| (JOHNSTON, 2020) | | | | | | | | |
| (KIM et al., 2023) | | | | | | | | |
| (KUHN; MALEKI, 2017) | | | | | | | | |
| (LAURSEN; NIELSEN; DYREBORG, 2021) | | | | | | | | |
| (LYKOURENTZOU; ROBERT JR.; BARLATIER, 2021) | | | | | | | | |
| (MA; KHANSA; KIM, 2018) | | | | | | | | |
| (OPPENLAENDER et al., 2020) | | | | | | | | |
| (ROSIN, 2021) | | | | | | | | |
| (SCHÖRPF et al., 2017) | | | | | | | | |
| (SEPPÄNEN et al., 2021) | | | | | | | | |
| (SHEVCHUK; STREBKOV, 2023) | | | | | | | | |
| (SHEVCHUK; STREBKOV; TYULYUPO, 2021) | | | | | | | | |
| (WANG et al., 2023) | | | | | | | | |
| (WONG et al., 2021) | | | | | | | | |

Figure 4. Classification of the 25 articles according to whether they touch (green) or not (red) the eight factors (Fn) related to the power dynamics, precarity, and tensions between crowdworkers and OLPs (A4).

As Figures 4 and 5 reveal, overall, the research papers strongly focus on autonomy, control, and unfair treatment, indicating these are key concerns in the crowdworking research field. However, it also suggests that there might be a potential need for more research in understanding the collective aspects of crowdwork, such as collective actions for better conditions and the impacts of reputation systems and algorithmic biases on crowdworkers. The subsequent subsections provide a detailed analysis of the key findings and themes identified in the selected articles.

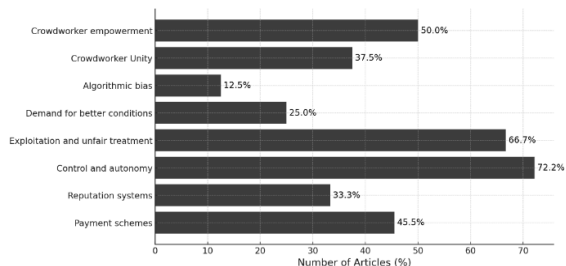

Figure 5. Percentage of the 25 selected articles that mention each of the eight factors (F1 to F8) underlying the aspects of power dynamics, precarity, and tensions between OLPs and workers (A4)

## 5.1 Payment Schemes

The schemes and timeliness of payments to online workers can significantly influence employee motivation, satisfaction, and overall engagement with an OLP (Zhang and Van der Schaar, 2012). While increasing pay doesn't necessarily ensure a consistent improvement in online work quality, it attracts more workers faster (Mason and Watts, 2010; Rogstadius et al., 2011). Furthermore, when extrinsic (e.g., payments) and intrinsic motivators are synergistically combined, a higher level of employee satisfaction and performance can be expected (Amabile, 1993). A payment system that is fair, transparent, and does not delay can foster workers' trust and long-term commitment to OLPs and recruiters (Rogstadius et al., 2011).

As mentioned by almost half of the articles, the role of payment schemes and their impact on the relationship between crowdworkers and online OLPs seems to be a noticeable factor from the research perspective. Possible reasons for this number not being higher, since this is an important aspect for workers, may include the fact that payment schemes in crowdworking OLPs, particularly with macrotasks, can vary widely and be complex, possibly making them a challenging subject for a comprehensive study. There might also be a lack of transparency from OLPs about their payment structures, posing a challenge for researchers. However, given the direct impact of payment schemes and mechanisms on the well-being of crowdworkers, and this being a factor linked to crowdworkers' adherence to SMECs (Wang et al., 2019; 2023), this could be a better and more deeply explored aspect in the future.

## 5.2 Trust and reputation systems

Trust and reputation systems serve as mechanisms to build trust between actual strangers in digital marketplaces where direct interaction is limited. Potential recruiters assess crowdworkers based on their profiles, particularly their ratings on an OLP. Crowdwork rating systems are thus at the core of the control over workers exerted by both recruiters and OLPs in this triangular relationship (Barnes, Green and Hoyos, 2015; Blair, 2001; Schörpf et al., 2017).

A third of the articles (33.3%) selected by the review devoted attention to this important topic that often directly influences the ability of workers to secure work, potentially better wages, and more flexibility of choice, impacting workers' livelihoods, satisfaction, and career progression.

However, how ratings are calculated, the potential for bias and the impact of negative reviews are critical issues that seem to be insufficiently investigated, as well as the impacts of these systems on workers' psychological well-being. The stress of maintaining high ratings, the social dynamics of feedback systems, and how these systems can be made more equitable and less prone to abuse seem to be opportunities to be further explored.

## 5.3 Control and autonomy

OLPs exert significant control over interactions and dependency management among platform participants (Schmidt, 2017), enforcing performance monitoring through various rules, policies, and standards (Deng, Joshi, and Galliers, 2016; Gandini, 2019). The Upwork platform, for instance, employs electronic monitoring, offering hourly workers a guarantee of earnings if they consent to periodic desktop screenshots and keystroke recording. This data and activity ratings are shared with clients (Kuhn and Maleki, 2017). While this system can mitigate the risk of nonpayment by enticing workers to compromise their privacy, it profoundly impacts their working conditions (Kaplan, 2016).

OLPs also typically centralize task decomposition, subdividing larger tasks into smaller, executable components (Khan et al., 2019; Lykourentzou et al., 2019) aiming at task quality and the ability to engage and control a broader spectrum of workers (Retelny, Bernstein, and Valentine, 2017).

Conversely, worker autonomy—particularly over task selection and execution—is a well-established determinant of job satisfaction and motivation, including the context of crowdworking (Baard, Deci, and Ryan, 2004; Ghezzi et al., 2018). The flexibility to select tasks, schedule work, interact with requesters, and maintain a degree of control is crucial in distinguishing between fulfilling and disheartening work experiences, especially since traditional supervisory frameworks are diminished or restructured in crowdworking scenarios.

Numerous articles explore the dimensions of control and autonomy, acknowledging the significant influence of OLPs in dividing and allocating tasks, reflecting their conventional function in structuring crowdwork. However, the effects of Small and Medium-sized Enterprise Contractors' enhanced autonomy and their proficiency in decomposing and overseeing complex tasks on worker satisfaction and motivation remain unclear. Similarly, how OLPs will adapt to this evolving landscape is yet to be comprehended.

## 5.4 Exploitation and unfair treatment

This factor is the second most mentioned among the articles selected and is a critical area of concern in the context of crowdworking and OLPs. Crowdwork notably lacks the traditional safeguards that protect workers in conventional employment (Gillespie, 2010). This absence of regulation and security can lead to situations where crowdworkers are vulnerable to exploitation or unfair treatment. Crowdworkers typically face irregular work hours, unpredictable income, and a lack of benefits such as health insurance, paid leave, or retirement plans. In the case of crowdfarms workers, for instance, workers experience increased communication costs, stress levels, and work schedules that resemble the 996 working hour system (Wang et al., 2020).

The legal status of crowdworkers is frequently ambiguous, raising questions about whether they are employees, independent contractors, or fall into a distinct category altogether. This lack of clarity introduces both ethical and legal complexities, potentially depriving workers of the protections and rights usually granted to traditional employees. Instead, they must depend exclusively on the policies and rules of OLPs, which often lack transparency and impartiality. In the context of SMECs, the insertion of an additional intermediary layer can exacerbate existing legal ambiguities or voids, potentially complicating the situation further. Additionally, the global dimension of crowdworking amplifies these challenges, allowing OLPs to capitalize on the increased vulnerability of workers in certain locales.

## 5.5 Demand for better conditions

Collective action and worker organizing can be powerful tools for workers to voice their concerns, negotiate better terms, and ensure fair practices. Trade unions or workers' organizations play this role in traditional employment sectors. However, in the decentralized world of crowdworking, organizing can be challenging, given the distributed nature of the workforce.

One-fourth of the papers addressing this topic suggest that the collective actions and demands of crowdworkers might not be as visible or well-documented as other aspects, making it harder for researchers to study them. The dispersed, individualized nature of crowdworking might lead to fewer collective, organized movements (Alacovska, Bucher and Fieseler, 2024; Johnston, 2020; Liu and Wang, 2022; Wood, Lehdonvirta and Graham, 2018), which in turn may result in less academic attention. We suspect that there might also be a bias in the academic community towards studying phenomena that are more easily quantifiable or align with OLP providers' interests rather than worker advocacy. And SMECs, as already established, situated, and recognized organizations, may eventually play an important role in collective action.

## 5.6 Algorithmic bias

Recruiter ratings to crowdworkers are shaped by algorithms into reputation scores (Seppänen et al., 2021; Wood et al., 2019). Trust and reputation systems play a well-known and fundamental role in online platforms, especially in the realm of crowdworking. Algorithms are also employed to match and form teams (Basu et al., 2014). So, it is surprising that only three papers directly touch on this topic, which could imply that while recognized as an issue, it may not be as extensively explored as other topics in the context of OLP crowdworking.

The topic of algorithmic bias in crowdworking might be still emerging. As awareness of the implications of AI and algorithms grows, this could become a more prominent research area. Research in this area also requires a deep understanding of both machine learning algorithms and the specific ways they interact with labor dynamics, which might be a barrier to some researchers.

Algorithms can inadvertently and intentionally introduce, perpetuate, or amplify biases, leading to unfair or discriminatory outcomes for certain crowdworker populations. However, the extent to which the phenomenon of SMEC is a defense or reaction against algorithmic effects is still unknown.

## 5.7 Crowdworker unity

Crowdworker unity and solidarity were mentioned by 37.5% of the articles, indicating a relatively moderate interest. Historically, gig workers have been viewed as isolated, but there's an increasing awareness of the potential for solidarity and collective bargaining, even in such dispersed work environments (Hau and Savage, 2023; Liu and Wang, 2022; Wood, Lehdonvirta and Graham, 2018; Woodside, Vinodrai and Moos, 2021).

Technology, while an enabler of the gig economy, presents both opportunities and challenges for worker organization (De La Torre-López,

Ramírez and Romero, 2023; Lykourentzou, Robert JR and Barlatier, 2022; Zhou and Pun, 2022). The unique nature of this technology-mediated form of work – where workers are often isolated and compete against each other for tasks (Soriano, 2021) – makes it a compelling area for research. Studies might explore how solidarity can be fostered in an environment typically characterized by individualized, remote work.

## 5.8 Crowdworker empowerment

Crowdworker empowerment and agency is about giving them more control, voice, and influence in OLPs' decision-making processes (Deng, Joshi and Galliers, 2016; Lykourentzou et al., 2019), especially in areas that directly affect their work, like, for instance, workflow definition and management (Retelny, Bernstein and Valentine, 2017). This can lead to more equitable OLP policies, improved worker satisfaction, and improved work outcomes.

Half of the articles mentioning this factor indicate academic interest in the agency of crowdworkers in shaping OLP policies and practices. This may indicate an evolving concern about exploitation and unfair practices, turning worker empowerment into a crucial area to be explored. How crowdworkers can assert their rights and influence the terms of their engagement is central to discussions about the future of fair work in OLPs. And as Wang et al. (2023) indicate, a search for self-empowerment, or at least protection, is underneath the organization of workers around SMECs like the Chinese crowdfarms.

# 6 LIMITATIONS OF THE STUDY

This review, while comprehensive, acknowledges certain limitations. The scope of literature, though broad, may not capture all emerging research within the SMEC and OLP domains. Analytical depth was sought, yet further exploration into the nuances of how identified factors influence SMECs could enrich understanding. Methodological transparency has been a priority; however, deeper justification for selection criteria could enhance rigor. The review strives for a balanced perspective, yet engagement with contradictory evidence could be strengthened. The limitations identified in this review set the stage for future research.

# 7 FUTURE RESEARCH DIRECTIONS

The review has unveiled numerous avenues for further research, highlighting critical questions that future studies could address to deepen our understanding of SMECs and their evolving role. These questions include:

- **Working Conditions**: How do SMECs influence worker income, job security, and overall well-being compared to traditional OLP work? What authentic pathways do SMECs create toward improved working conditions and more balanced power dynamics for crowdworkers?
- **Talent Acquisition**: How can SMECs attract and retain skilled workers while maintaining a healthy internal structure, remaining competitive, and managing potential declines in platform payments? How the competition for talent may influence the relationship between SMECs and OLPs?
- **Legal and Regulatory Frameworks**: How can legal frameworks evolve to recognize and effectively regulate the unique characteristics of SMECs? Additionally, what regulatory policies are necessary to foster fair competition among SMECs and between these entities and individual workers and ensure their protection?
- **OLP Adaptation**: In what ways might OLPs adapt to the emergence of SMECs? Will they tolerate, integrate, co-opt these models, or promote more competition?
- **Scalability and Technological Advancements**: What technological advancements or organizational structures could support scalability within SMECs? Can we anticipate the formation of cooperative networks among SMECs, and if so, what might these arrangements look like?

Addressing these questions could significantly contribute to the knowledge of OLPs, offering valuable insights for academics, practitioners, policymakers, and crowdworkers.

# 8 CONCLUSIONS

The evolution of crowdworking towards autonomous, enterprise-like models marks a significant shift from traditional, platform-controlled work. Our review focuses on the platform-centric perspective in crowdworking, addressing RQ1 by examining how crowdwork aspects integral to SMECs, such as task decomposition and team coordination, are discussed in the literature. We uncover a gap in understanding the power dynamics between crowdworkers and OLPs (RQ2), particularly in areas like payment, autonomy, and algorithmic bias. Identifying future research avenues, including exploring exploitation and stakeholder balance (RQ3), underscores the need for a holistic approach. This review contributes to a more comprehensive understanding of crowdworking, advocating for theoretical and practical advancements that prioritize the well-being and empowerment of crowdworkers.